\documentclass[preprintnumbers,amsmath,amssymbm,prd]{revtex4}
\usepackage{epsfig}
\usepackage{graphicx}
\usepackage{amssymb}

\begin{document}
%\title{Marginally stable exotic compact objects}
\title{Onset of superradiant instabilities in rotating spacetimes of exotic compact objects}
\author{Shahar Hod}
\affiliation{The Ruppin Academic Center, Emeq Hefer 40250, Israel}
\affiliation{ } \affiliation{The Hadassah Institute, Jerusalem
91010, Israel}
\date{\today}

\begin{abstract}
\ \ \ Exotic compact objects, horizonless spacetimes with reflective
properties, have intriguingly been suggested by some quantum-gravity
models as alternatives to classical black-hole spacetimes. A
remarkable feature of spinning horizonless compact objects with
reflective boundary conditions is the existence of a {\it discrete}
set of critical surface radii, $\{r_{\text{c}}({\bar
a};n)\}^{n=\infty}_{n=1}$, which can support spatially regular
static ({\it marginally-stable}) scalar field configurations (here
${\bar a}\equiv J/M^2$ is the dimensionless angular momentum of the
exotic compact object). Interestingly, the outermost critical radius
$r^{\text{max}}_{\text{c}}\equiv \text{max}_n\{r_{\text{c}}({\bar
a};n)\}$ marks the boundary between stable and unstable exotic
compact objects: spinning objects whose reflecting surfaces are
situated in the region $r_{\text{c}}>r^{\text{max}}_{\text{c}}({\bar
a})$ are stable, whereas spinning objects whose reflecting surfaces
are situated in the region
$r_{\text{c}}<r^{\text{max}}_{\text{c}}({\bar a})$ are
superradiantly unstable to scalar perturbation modes. In the present
paper we use analytical techniques in order to explore the physical
properties of the critical (marginally-stable) spinning exotic
compact objects. In particular, we derive a remarkably compact {\it
analytical} formula for the discrete spectrum
$\{r^{\text{max}}_{\text{c}}({\bar a})\}$ of critical radii which
characterize the marginally-stable exotic compact objects. We
explicitly demonstrate that the analytically derived resonance
spectrum agrees remarkably well with numerical results that recently
appeared in the physics literature.
\end{abstract}
\bigskip
\maketitle

%]

\section{Introduction}

Black holes are certainly the most important prediction of general
relativity, Einstein's classical theory of gravity. These
fundamental objects are characterized by curved spacetime geometries
with absorbing boundary conditions (event horizons). Intriguingly,
however, some researchers (see
\cite{eco1,eco2,eco3,eco4,eco5,eco6,eco7,eco8,eco9,eco10,eco11,eco12,Pan}
and references therein) have recently argued that black-hole
horizons may be quantum-mechanically unstable. Horizonless compact
objects has therefore been proposed
\cite{eco1,eco2,eco3,eco4,eco5,eco6,eco7,eco8,eco9,eco10,eco11,eco12,Pan}
as exotic quantum-gravity alternatives to the familiar (classical)
black-hole spacetimes .

It is certainly of physical importance to explore the (in)stability
properties of these horizonless exotic compact objects. In a very
important work, Maggio, Pani, and Ferrari \cite{Pan} have recently
demonstrated numerically that, due to the physical mechanism of
superradiant amplification \cite{Frid,Bri}, horizonless spinning
compact objects with reflective boundary conditions and ergoregions
may become unstable to scalar perturbation modes \cite{Notebi}.
Intriguingly, the results presented in \cite{Pan} have revealed the
important fact that the ergoregion instability shuts down if the
quantum reflective surface of the spinning exotic compact object is
located far enough from the would-be classical black-hole horizon.
This interesting and highly important numerical finding implies, in
particular, that there exists a unique family of critical ({\it
marginally-stable}) exotic compact objects which determine the
boundary between stable and unstable horizonless spinning
configurations.

In the present paper we shall use {\it analytical} techniques in
order to explore the physical properties of the critical
(marginally-stable) exotic compact objects. In particular, below we
shall explicitly prove that horizonless spinning compact objects
with reflective boundary conditions are characterized by the
existence of a discrete set $\{r_{\text{c}}({\bar
a};n)\}^{n=\infty}_{n=1}$ of critical surface radii that can support
spatially regular {\it static} (marginally-stable) massless scalar
field configurations \cite{Notemsm,Hodrc,Herkr} [Here ${\bar a}$ is
the dimensionless angular momentum of the exotic compact object, see
Eq. (\ref{Eq12}) below].

It should be emphasized that the physical significance of the
outermost (largest) radius $r^{\text{max}}_{\text{c}}({\bar
a})\equiv \text{max}_n\{r_{\text{c}}({\bar a};n)\}$ stems from the
fact that, for a given value of the dimensionless rotation parameter
${\bar a}$, this unique critical radius marks the boundary between
stable and unstable spinning exotic compact configurations. In
particular, horizonless compact objects of rotation parameter ${\bar
a}$ whose reflecting surfaces are characterized by the inequality
$r_{\text{c}}<r^{\text{max}}_{\text{c}}({\bar a})$ are
superradiantly unstable to scalar perturbation modes, whereas
horizonless compact objects whose reflecting surfaces are
characterized by the inequality
$r_{\text{c}}>r^{\text{max}}_{\text{c}}({\bar a})$ are stable.

The main goal of the present paper is to determine {\it
analytically} the discrete spectrum $\{r_{\text{c}}({\bar
a};n)\}^{n=\infty}_{n=1}$ of radii which characterize the
horizonless spinning exotic compact objects that can support the
spatially regular static (marginally-stable) scalar field
configurations. In particular, below we shall derive a remarkably
compact analytical formula for the critical (outermost) radii
$\{r^{\text{max}}_{\text{c}}({\bar a})\}$ of the exotic compact
objects that mark the boundary between stable and unstable
horizonless spinning configurations.

\section{Description of the system}

We shall analyze the physical properties of horizonless spinning
compact objects with reflective surfaces which are linearly coupled
to massless scalar fields. Following the interesting work of Maggio,
Pani, and Ferrari \cite{Pan} (see also
\cite{eco1,eco2,eco3,eco4,eco5,eco6,eco7,eco8,eco9,eco10,eco11,eco12}),
we shall study exotic compact objects which are characterized by
curved geometries that modify the spinning Kerr spacetime only at
some microscopic scale around the would-be classical horizon. In
particular, we shall assume, as in \cite{Pan}, that the Kerr line
element \cite{Chan,Noteun,Notebl,Noteman}
\begin{eqnarray}\label{Eq1}
ds^2=-{{\Delta}\over{\rho^2}}(dt-a\sin^2\theta
d\phi)^2+{{\rho^2}\over{\Delta}}dr^2+\rho^2
d\theta^2+{{\sin^2\theta}\over{\rho^2}}\big[a
dt-(r^2+a^2)d\phi\big]^2\ \ \ \text{for}\ \ \ \ r>r_{\text{c}}\  ,
\end{eqnarray}
with $\Delta\equiv r^2-2Mr+a^2$ and $\rho^2\equiv
r^2+a^2\cos^2\theta$, describes the external spacetime of the
spinning exotic compact object of mass $M$, angular momentum $J=Ma$,
and radius $r_{\text{c}}$. In addition, below we shall assume that
the radius of the exotic compact object is characterized by the
relation \cite{Notesk,kr1,kr2,kr3}
\begin{equation}\label{Eq2}
z_{\text{c}}\equiv {{r_{\text{c}}-r_+}\over{r_+}}\ll1\  ,
\end{equation}
where
\begin{equation}\label{Eq3}
r_{\pm}=M+(M^2-a^2)^{1/2}\
\end{equation}
are the radii of the would-be classical horizons. The
small-$z_{\text{c}}$ regime (\ref{Eq2}) corresponds to the
physically interesting family of horizonless exotic compact objects
whose quantum reflective surfaces are located a microscopic distance
above the would-be classical black-hole horizons \cite{Pan} (see
also
\cite{eco1,eco2,eco3,eco4,eco5,eco6,eco7,eco8,eco9,eco10,eco11,eco12}).

The dynamics of the linearized massless scalar field in the spinning
spacetime of the exotic compact object is governed by the
Klein-Gordon wave equation \cite{Teuk,Stro}
\begin{equation}\label{Eq4}
\nabla^\nu\nabla_{\nu}\Psi=0\  .
\end{equation}
Substituting into (\ref{Eq4}) the metric components of the curved
line element (\ref{Eq1}), which characterizes the exterior spacetime
of the spinning exotic compact object, and using the scalar field
decomposition \cite{Teuk,Stro,Notedec}
\begin{equation}\label{Eq5}
\Psi=\sum_{l,m}e^{im\phi}{S_{lm}}(\theta;a\omega){R_{lm}}(r;M,a,\omega)e^{-i\omega
t}\  ,
\end{equation}
one finds that the radial part $R_{lm}(r;M,a,\omega)$ of the scalar
eigenfunction is determined by the ordinary differential equation
\cite{Teuk,Stro}
\begin{equation}\label{Eq6}
\Delta{{d}
\over{dr}}\Big(\Delta{{dR_{lm}}\over{dr}}\Big)+\Big\{[\omega(r^2+a^2)-ma]^2
+\Delta(2ma\omega-K_{lm})\Big\}R_{lm}=0\ .
\end{equation}
Here the frequency-dependent parameter $K_{lm}(a\omega)$ is the
characteristic eigenvalue of the spatially regular angular
eigenfunction $S_{lm}(\theta;a\omega)$, which is determined by the
familiar spheroidal differential equation
\cite{Heun,Fiz1,Teuk,Abram,Stro,Hodasy,Hodpp}
\begin{eqnarray}\label{Eq7}
{1\over {\sin\theta}}{{d}\over{\theta}}\Big(\sin\theta {{d
S_{lm}}\over{d\theta}}\Big) +\Big(K_{lm}-a^2\omega^2
+a^2\omega^2\cos^2\theta-{{m^2}\over{\sin^2\theta}}\Big)S_{lm}=0\ .
\end{eqnarray}
For later purposes we note that, in the small frequency
$a\omega\ll1$ regime, the angular eigenvalues of the spheroidal
scalar eigenfunctions can be expanded in the from
\begin{equation}\label{Eq8}
K_{lm}-a^2\omega^2=l(l+1)+\sum_{k=1}^{\infty}c_k (a\omega)^{2k}\ ,
\end{equation}
where the frequency-independent coefficients $\{c_k(l,m)\}$ are
given in \cite{Abram}.

Following the physically interesting model of the spinning exotic
objects studied numerically in \cite{Pan}, we shall assume that the
horizonless compact objects are characterized by (Dirichlet or
Neumann) reflecting boundary conditions \cite{Noteby}:
\begin{equation}\label{Eq9}
\begin{cases}
R(r=r_{\text{c}})=0 &\ \ \ \ \text{Dirichlet B. C.}\ ; \\
dR(r=r_{\text{c}})/dr=0 &\ \ \ \ \text{Neumann B. C.}\ \ .
\end{cases}
\end{equation}
In addition, the marginally-stable (static) scalar modes that we
shall analyze in the present paper are characterized by spatially
regular (asymptotically decaying) eigenfunctions:
\begin{equation}\label{Eq10}
R(r\to\infty)\to 0\  .
\end{equation}

\section{The resonance conditions of the marginally-stable spinning
exotic compact objects}

The composed horizonless-spinning-object-massless-scalar-field
system is characterized by the existence of a unique family of
marginally-stable (static) resonances which mark the onset of
superradiant instabilities in the curved spinning spacetime. These
physically interesting critical field modes are characterized by the
simple property
\begin{equation}\label{Eq11}
\omega=0\  .
\end{equation}

The set of equations (\ref{Eq6})-(\ref{Eq11}) determines two {\it
discrete} spectra of radii, $\{r^{\text{Dirichlet}}_{\text{c}}({\bar
a},l,m;n)\}$ and $\{r^{\text{Neumann}}_{\text{c}}({\bar
a},l,m;n)\}$, which, for a given value
\begin{equation}\label{Eq12}
{\bar a}\equiv {{J}\over{M^2}}\
\end{equation}
of the dimensionless angular momentum parameter, characterize the
critical (marginally-stable) spinning exotic compact objects.
Interestingly, as we shall explicitly show in the present section,
the characteristic radial equation (\ref{Eq6}) of the massless
scalar fields in the curved spacetimes of the spinning exotic
compact objects is amenable to an {\it analytical} treatment for the
marginally-stable static modes.

Substituting into Eq. (\ref{Eq6}) $\omega=0$ and
\begin{equation}\label{Eq13}
R(x)=x^{-i\alpha}(1-x)^{l+1}F(x)\  ,
\end{equation}
where
\begin{equation}\label{Eq14}
x\equiv {{r-r_+}\over{r-r_-}}\ \ \ \ ; \ \ \ \
\alpha\equiv{{ma}\over{r_+-r_-}}\  ,
\end{equation}
one obtains the characteristic radial equation \cite{Noteklm0}
\begin{equation}\label{Eq15}
x(1-x){{d^2F}\over{dx^2}}+\{(1-2i\alpha)-[1+2(l+1)-2i\alpha]x\}{{dF}\over{dx}}-
[(l+1)^2-2i\alpha(l+1)]F=0\  .
\end{equation}
The general mathematical solution of the radial differential
equation (\ref{Eq15}) is given by \cite{Abram,Morse}
\begin{eqnarray}\label{Eq16}
F(x)=A\cdot{_2F_1}(l+1-2i\alpha,l+1;2l+2;1-x)+B\cdot
(1-x)^{-2l-1}{_2F_1}(-l-2i\alpha,-l;-2l;1-x)\ ,
%F(x)=A\cdot{_2F_1}(l+1-2i\alpha,l+1;1-2i\alpha;x)+B\cdot
%x^{2i\alpha}{_2F_1}(l+1,l+1+2i\alpha;1+2i\alpha;x)\ ,
\end{eqnarray}
where $_2F_1(a,b;c;z)$ is the hypergeometric function and $\{A,B\}$
are normalization constants. Substituting (\ref{Eq16}) into
(\ref{Eq13}), one obtains the expression
\begin{eqnarray}\label{Eq17}
R(x)=x^{-i\alpha}[A\cdot(1-x)^{l+1}{_2F_1}(l+1-2i\alpha,l+1;2l+2;1-x)+B\cdot
(1-x)^{-l}{_2F_1}(-l-2i\alpha,-l;-2l;1-x)]\
%R(x)=(1-x)^{l+1}\cdot[A\cdot
%x^{-i\alpha}{_2F_1}(l+1-2i\alpha,l+1;1-2i\alpha;x)+B\cdot
%x^{i\alpha}{_2F_1}(l+1,l+1+2i\alpha;1+2i\alpha;x)]\ .
\end{eqnarray}
for the radial scalar eigenfunction. Using the characteristic
property (see Eq. 15.1.1 of \cite{Abram})
\begin{eqnarray}\label{Eq18}
{_2F_1}(a,b;c;z\to0)\to1\
\end{eqnarray}
of the hypergeometric function, one finds from (\ref{Eq17}) the
asymptotic $r\to\infty$ [$x\to1^-$, see Eq. (\ref{Eq14})] behavior
\begin{eqnarray}\label{Eq19}
R(x\to1)=A\cdot(1-x)^{l+1}+B\cdot (1-x)^{-l}\
\end{eqnarray}
of the radial eigenfunction. A physically acceptable (finite)
solution at spatial infinity ($r\to\infty$, or equivalently $x\to
1^-$) requires
\begin{equation}\label{Eq20}
B=0\  .
\end{equation}

We therefore conclude that the marginally-stable (static) resonances
of the massless scalar fields in the curved spacetimes of the
horizonless spinning exotic compact objects are characterized by the
radial eigenfunction
\begin{equation}\label{Eq21}
R(x)=A\cdot
x^{-i\alpha}(1-x)^{l+1}{_2F_1}(l+1-2i\alpha,l+1;2l+2;1-x)\  .
\end{equation}
Taking cognizance of the boundary conditions (\ref{Eq9}), which
characterize the horizonless reflecting compact objects, one deduces
that the compact resonance equations
\begin{equation}\label{Eq22}
{_2F_1}(l+1-2i\alpha,l+1;2l+2;1-x_{\text{c}})=0\ \ \ \ \ \text{for}\
\ \ \ \ \text{Dirichlet B. C.}\
\end{equation}
and
\begin{equation}\label{Eq23}
{{d}\over{dx}}[x^{-i\alpha}(1-x)^{l+1}{_2F_1}(l+1-2i\alpha,l+1;2l+2;1-x)]_{x=x_{\text{c}}}=0\
\ \ \ \ \text{for}\ \ \ \ \ \text{Neumann B. C.}\
\end{equation}
determine the {\it discrete} spectra of dimensionless critical radii
$\{x_{\text{c}}({\bar a},l,m;n)\}$ which characterize the
marginally-stable exotic compact objects.

\section{The discrete resonance spectra of the marginally-stable spinning exotic compact objects}

The analytically derived resonance equations (\ref{Eq22}) and
(\ref{Eq23}), which characterize the unique families of horizonless
exotic compact objects that can support the static ({\it
marginally-stable}) massless scalar field configurations, can easily
be solved numerically. Interestingly, one finds that, for given
physical parameters $\{{\bar a},l,m\}$ of the composed
spinning-exotic-compact-object-massless-scalar-field system, there
exists a {\it discrete} set of critical radii,
\begin{equation}\label{Eq24}
\cdots r_{\text{c}}(n=3)<r_{\text{c}}(n=2)<r_{\text{c}}(n=1)\equiv
r^{\text{max}}_{\text{c}}({\bar a},l,m)\ ,
\end{equation}
which can support the spatially regular static (marginally-stable)
scalar field resonances.

In Table \ref{Table1} we display, for various values of the
dimensionless angular momentum parameter ${\bar a}$, the largest
({\it outermost}) dimensionless radii \cite{Notezx}
\begin{equation}\label{Eq25}
z^{\text{max}}_{\text{c}}({\bar a},l,m)\equiv
{{r^{\text{max}}_{\text{c}}-r_+}\over{r_+}}\
\end{equation}
of the horizonless spinning compact objects that can support the
marginally-stable massless scalar field configurations. From the
data presented in Table \ref{Table1} one learns that, for fixed
values of the scalar angular harmonic indices, the critical radii
$z^{\text{max}}_{\text{c}}({\bar a})$ [and thus also
$r^{\text{max}}_{\text{c}}({\bar a})$] which characterize the {\it
marginally-stable} spinning exotic compact objects are a
monotonically increasing function of the dimensionless angular
momentum parameter ${\bar a}$.

It is worth emphasizing again that the physical significance of the
critical reflecting radius $r^{\text{max}}_{\text{c}}({\bar a}) $
stems from the fact that, for a given value of the dimensionless
angular momentum parameter ${\bar a}$, this supporting radius
corresponds to a {\it marginally-stable} spinning object which marks
the onset of superradiant instabilities in the composed
exotic-compact-object-massless-scalar-field system. In particular,
as nicely demonstrated numerically in \cite{Pan}, spinning compact
objects of dimensionless angular momentum ${\bar a}$ whose
reflecting surfaces are located in the region
$r_{\text{c}}>r^{\text{max}}_{\text{c}}({\bar a})$ are stable,
whereas spinning objects whose reflecting surfaces are located in
the region $r_{\text{c}}<r^{\text{max}}_{\text{c}}({\bar a})$ are
superradiantly unstable to scalar perturbation modes.

\begin{table}[htbp]
\centering
\begin{tabular}{|c|c|c|c|c|c|c|c|c|}
\hline \text{Boundary conditions} & \
$z^{\text{max}}_{\text{c}}({\bar a}=0.3)$ \ \ & \
$z^{\text{max}}_{\text{c}}({\bar a}=0.5)$ \ \ & \
$z^{\text{max}}_{\text{c}}({\bar a}=0.7)$\ \ & \
$z^{\text{max}}_{\text{c}}({\bar a}=0.9)$\ \ & \
$z^{\text{max}}_{\text{c}}({\bar a}=0.99)$\
\ & \ $z^{\text{max}}_{\text{c}}({\bar a}=0.999)$\ \ \\
\hline \ \ \text{Dirichlet}\ \ \ &\ \ $2.960\times10^{-10}$\ \ &\ \
$2.842\times10^{-6}$\ \ &\ \ $2.783\times10^{-4}$\ \ &\ \
$1.007\times10^{-2}$\ \ &\ \
$9.625\times10^{-2}$\ \ &\ \ $1.730\times10^{-1}$\ \ \\
\ \ \text{Neumann}\ \ \ &\ \ $6.455\times10^{-6}$\ \ &\ \
$6.662\times10^{-4}$\ \ &\ \ $7.417\times10^{-3}$\ \ &\ \
$5.432\times10^{-2}$\ \ &\ \
$2.105\times10^{-1}$\ \ &\ \ $3.078\times10^{-1}$\ \ \\
\hline
\end{tabular}
\caption{Marginally-stable spinning exotic compact objects. We
display, for various values of the dimensionless angular momentum
parameter ${\bar a}$, the largest dimensionless radii
$z^{\text{max}}_{\text{c}}({\bar a})\equiv
(r^{\text{max}}_{\text{c}}-r_+)/r_+$ [see Eqs. (\ref{Eq24}) and
(\ref{Eq25})] of the horizonless reflecting compact objects that can
support the spatially regular static scalar field configurations
with $l=m=1$. One finds that the critical radii
$\{z^{\text{max}}_{\text{c}}({\bar a})\}$, which characterize the
marginally-stable spinning exotic compact objects, are a
monotonically increasing function of the dimensionless angular
momentum parameter ${\bar a}$.} \label{Table1}
\end{table}

In Table \ref{Table2} we display, for various equatorial ($l=m$)
massless scalar field modes, the critical (largest) dimensionless
radii $z^{\text{max}}_{\text{c}}(l)$ [see Eq. (\ref{Eq25})] of the
horizonless spinning compact objects that can support the static
(marginally-stable) scalar field resonances. From the data presented
in Table \ref{Table2} one learns that, for a fixed value of the
dimensionless angular momentum parameter ${\bar a}$, the
dimensionless critical radii $z^{\text{max}}_{\text{c}}(l)$ [and
thus also $r^{\text{max}}_{\text{c}}(l)$] which characterize the
{\it marginally-stable} spinning exotic compact objects are a
monotonically increasing function of the scalar harmonic index $l$.

\begin{table}[htbp]
\centering
\begin{tabular}{|c|c|c|c|c|c|c|c|c|}
\hline \text{Boundary conditions} & \
$z^{\text{max}}_{\text{c}}(l=1)$ \ \ & \
$z^{\text{max}}_{\text{c}}(l=2)$ \ \ & \
$z^{\text{max}}_{\text{c}}(l=3)$\ \ & \
$z^{\text{max}}_{\text{c}}(l=4)$\ \ & \
$z^{\text{max}}_{\text{c}}(l=5)$\
\ & \ $z^{\text{max}}_{\text{c}}(l=6)$\ \ \\
\hline \ \ \text{Dirichlet}\ \ \ &\ \ $2.842\times10^{-6}$\ \ &\ \
$3.689\times10^{-4}$\ \ &\ \ $1.788\times10^{-3}$\ \ &\ \
$3.948\times10^{-3}$\ \ &\ \
$6.395\times10^{-3}$\ \ &\ \ $8.870\times10^{-3}$\ \ \\
\ \ \text{Neumann}\ \ \ &\ \ $6.662\times10^{-4}$\ \ &\ \
$6.141\times10^{-3}$\ \ &\ \ $1.268\times10^{-2}$\ \ &\ \
$1.832\times10^{-2}$\ \ &\ \
$2.297\times10^{-2}$\ \ &\ \ $2.680\times10^{-2}$\ \ \\
\hline
\end{tabular}
\caption{Marginally-stable spinning exotic compact objects. We
present the largest dimensionless radii
$z^{\text{max}}_{\text{c}}(l)\equiv
(r^{\text{max}}_{\text{c}}-r_+)/r_+$ [see Eqs. (\ref{Eq24}) and
(\ref{Eq25})] of the horizonless reflecting compact objects with
dimensionless angular momentum parameter ${\bar a}=0.5$ that can
support static equatorial ($l=m$) scalar field configurations. One
finds that the critical radii $\{z^{\text{max}}_{\text{c}}(l)\}$,
which characterize the marginally-stable spinning exotic compact
objects, are a monotonically increasing function of the scalar
harmonic index $l$.} \label{Table2}
\end{table}

\section{Resonance spectra for highly-compact spinning exotic objects}

\subsection{An analytical treatment}

In the present section we shall explicitly prove that the resonance
conditions (\ref{Eq22}) and (\ref{Eq23}), which respectively
determine the dimensionless discrete radii
$\{x^{\text{Dirichlet}}_{\text{c}}({\bar a},l,m;n)\}$ and
$\{x^{\text{Neumann}}_{\text{c}}({\bar a},l,m;n)\}$ of the
marginally-stable horizonless spinning exotic compact objects, can
be solved {\it analytically} in the physically interesting regime
\begin{equation}\label{Eq26}
x_{\text{c}}\ll1\  .
\end{equation}
As emphasized above, the small-$x_{\text{c}}$ regime (\ref{Eq26})
corresponds to the physically interesting family of highly compact
exotic objects whose quantum reflective surfaces are located very
near the would-be classical black-hole horizons. It is worth
mentioning, in particular, that the strong inequality (\ref{Eq26})
characterizes the horizonless spinning exotic compact objects that
were recently studied numerically in the interesting work of Maggio,
Pani, and Ferrari \cite{Pan} (see also
\cite{eco1,eco2,eco3,eco4,eco5,eco6,eco7,eco8,eco9,eco10,eco11,eco12}).

Using Eq. 15.3.6 of \cite{Abram}, one can express the characteristic
radial scalar eigenfunction (\ref{Eq21}) in the form \cite{Notehg}
\begin{eqnarray}\label{Eq27}
R(x)&=&A\cdot{{(2l+1)!}\over{l!}}(1-x)^{l+1}\Big[{{\Gamma(2i\alpha)}\over{\Gamma(l+1+2i\alpha)}}
x^{-i\alpha}{_2F_1}(l+1-2i\alpha,l+1;1-2i\alpha;x)\nonumber \\&& +
{{\Gamma(-2i\alpha)}\over{\Gamma(l+1-2i\alpha)}}
x^{i\alpha}{_2F_1}(l+1+2i\alpha,l+1;1+2i\alpha;x)\Big]\ .
\end{eqnarray}
Taking cognizance of the characteristic asymptotic property
(\ref{Eq18}) of the hypergeometric functions, one finds from
(\ref{Eq27}) the small-$x$ behavior
\begin{eqnarray}\label{Eq28}
R(x\ll1)=A{{(2l+1)!}\over{l!}}(1-x)^{l+1}\Big[{{\Gamma(2i\alpha)}\over{\Gamma(l+1+2i\alpha)}}
x^{-i\alpha}+{{\Gamma(-2i\alpha)}\over{\Gamma(l+1-2i\alpha)}}x^{i\alpha}\Big]\cdot[1+O(x)]\
\end{eqnarray}
of the radial scalar eigenfunction.

Using the characteristic small-$x$ spatial behavior (\ref{Eq28}) of
the radial scalar eigenfunction, one can express the Dirichlet and
Neumann resonance equations (\ref{Eq22}) and (\ref{Eq23}) in the
remarkably compact form \cite{Notehg}
\begin{equation}\label{Eq29}
x^{2i\alpha}=\mp{{\Gamma(2i\alpha)\Gamma(l+1-2i\alpha)}\over{\Gamma(-2i\alpha)\Gamma(l+1+2i\alpha)}}\
,
\end{equation}
where the upper/lower signs in (\ref{Eq29}) refer respectively to
the reflecting Dirichlet/Neumann boundary conditions. From the
resonance conditions (\ref{Eq29}) one finally finds the compact
analytical formulas \cite{Notenn,Notegam1}
\begin{equation}\label{Eq30}
x^{\text{Dirichlet}}_{\text{c}}(n)={e^{-\pi
(n+{1\over2})/\alpha}}\Big[{{\Gamma(2i\alpha)\Gamma(l+1-2i\alpha)}\over{\Gamma(-2i\alpha)\Gamma(l+1+2i\alpha)}}\Big]
^{1/2i\alpha}\ \ \ ; \ \ \ n\in\mathbb{Z}
\end{equation}
and
\begin{equation}\label{Eq31}
x^{\text{Neumann}}_{\text{c}}(n)={e^{-\pi
n/\alpha}}\Big[{{\Gamma(2i\alpha)\Gamma(l+1-2i\alpha)}\over{\Gamma(-2i\alpha)\Gamma(l+1+2i\alpha)}}\Big]
^{1/2i\alpha}\ \ \ ; \ \ \ n\in\mathbb{Z}
\end{equation}
for the discrete families
$\{x^{\text{Dirichlet}}_{\text{c}}(n),x^{\text{Neumann}}_{\text{c}}(n)\}$
of dimensionless critical radii which characterize the horizonless
spinning compact objects that can support the spatially regular
static (marginally-stable) massless scalar field resonances.

\subsection{Numerical confirmation}

It is physically important to verify the validity of the
analytically derived resonance spectra (\ref{Eq30}) and (\ref{Eq31})
for the characteristic discrete radii of the highly compact
($x_{\text{c}}\ll1$) spinning exotic objects that can support the
static (marginally-stable) scalar field configurations. In Tables
\ref{Table3} and \ref{Table4} we display the dimensionless radii
$z^{\text{analytical}}_{\text{c}}(n)\equiv
[r_{\text{c}}(n)-r_+]/r_+$ \cite{Notezx} of the spinning exotic
compact objects with reflecting Dirichlet/Neumann boundary
conditions as calculated from the analytically derived resonance
spectra (\ref{Eq30}) and (\ref{Eq31}). For comparison, we also
display the corresponding radii $z^{\text{numerical}}_{\text{c}}(n)$
of the horizonless compact objects as obtained from a direct
numerical solution of the exact (analytically derived) resonance
conditions (\ref{Eq22}) and (\ref{Eq23}).

From the data presented in Tables \ref{Table3} and \ref{Table4} one
finds a very good agreement, especially in the physically
interesting regime $z_{\text{c}}\ll1$ of highly compact exotic
objects \cite{Notephy} [see Eq. (\ref{Eq2})], between the
approximated radii of the horizonless compact objects that can
support the static (marginally-stable) scalar resonances [as
calculated from the analytical formulas (\ref{Eq30}) and
(\ref{Eq31})] and the corresponding exact radii of the spinning
compact objects [as determined numerically from the characteristic
resonance conditions (\ref{Eq22}) and (\ref{Eq23})].

\begin{table}[htbp]
\centering
\begin{tabular}{|c|c|c|c|c|c|}
\hline \text{Formula} & \ $z^{\text{Dir}}_{\text{c}}(n=1)$\ \ & \
$z^{\text{Dir}}_{\text{c}}(n=2)$\ \ & \
$z^{\text{Dir}}_{\text{c}}(n=3)$\ \ & \ $z^{\text{Dir}}_{\text{c}}(n=4)$\ \ & \ $z^{\text{Dir}}_{\text{c}}(n=5)$\ \ \\
\hline \ {\text{Analytical}}\ [Eq. (\ref{Eq30})]\ \ &\
$9.947\times10^{-3}$\ \ &\ $4.671\times10^{-4}$\ \
&\ $2.226\times10^{-5}$\ \ &\ $1.061\times10^{-6}$\ \ &\ $5.061\times10^{-8}$\ \ \\
\ {\text{Numerical}}\ [Eq. (\ref{Eq22})]\ \ &\ $1.007\times10^{-2}$\
\ &\ $4.673\times10^{-4}$\ \ &\ $2.228\times10^{-5}$\ \ &\
$1.062\times10^{-6}$\ \
&\ $5.061\times10^{-8}$\ \ \\
\hline
\end{tabular}
\caption{Spinning exotic compact objects with reflective Dirichlet
boundary conditions. We present the analytically calculated discrete
set of dimensionless radii $z^{\text{analytical}}_{\text{c}}(n)$
which characterize the spinning compact objects that can support the
static (marginally-stable) massless scalar field configurations. We
also present the corresponding radii
$z^{\text{numerical}}_{\text{c}}(n)$ of the horizonless compact
objects as obtained from a direct numerical solution of the
characteristic resonance condition (\ref{Eq22}). The data presented
is for horizonless compact objects with dimensionless rotation
parameter ${\bar a}=0.9$ linearly coupled to a massless scalar field
mode with $l=m=1$. In the physically interesting regime
$z_{\text{c}}\ll1$ of highly compact exotic objects \cite{Notephy},
one finds a remarkably good agreement between the approximated radii
$\{z^{\text{analytical}}_{\text{c}}(n)\}$ of the compact objects
that can support the marginally-stable scalar resonances [as
calculated from the analytical resonance spectrum (\ref{Eq30})] and
the corresponding exact radii
$\{z^{\text{numerical}}_{\text{c}}(n)\}$ of the compact exotic
objects [as determined by a direct numerical solution of the
resonance equation (\ref{Eq22})].} \label{Table3}
\end{table}

\begin{table}[htbp]
\centering
\begin{tabular}{|c|c|c|c|c|c|}
\hline \text{Formula} & \ $z^{\text{Neu}}_{\text{c}}(n=1)$\ \ & \
$z^{\text{Neu}}_{\text{c}}(n=2)$\ \ & \
$z^{\text{Neu}}_{\text{c}}(n=3)$\ \ & \ $z^{\text{Neu}}_{\text{c}}(n=4)$\ \ & \ $z^{\text{Neu}}_{\text{c}}(n=5)$\ \ \\
\hline \ {\text{Analytical}}\ [Eq. (\ref{Eq31})]\ \ &\
$4.839\times10^{-2}$\ \ &\ $2.145\times10^{-3}$\ \
&\ $1.019\times10^{-4}$\ \ &\ $4.860\times10^{-6}$\ \ &\ $2.318\times10^{-7}$\ \ \\
\ {\text{Numerical}}\ [Eq. (\ref{Eq23})]\ \ &\ $5.432\times10^{-2}$\
\ &\ $2.153\times10^{-3}$\ \ &\ $1.020\times10^{-4}$\ \ &\
$4.860\times10^{-6}$\ \ &\ $2.318\times10^{-7}$\ \ \\
\hline
\end{tabular}
\caption{%Same as Table \ref{Table3}, but for spinning exotic compact
%objects with reflective Neumann boundary conditions.
Spinning exotic compact objects with reflective Neumann boundary
conditions. We present the analytically calculated discrete set of
dimensionless radii $z^{\text{analytical}}_{\text{c}}(n)$ which
characterize the spinning compact objects that can support the
static (marginally-stable) massless scalar field configurations. We
also present the corresponding radii
$z^{\text{numerical}}_{\text{c}}(n)$ of the horizonless compact
objects as obtained from a direct numerical solution of the
characteristic resonance condition (\ref{Eq23}). The data presented
is for horizonless compact objects with dimensionless rotation
parameter ${\bar a}=0.9$ and a massless scalar field mode with
$l=m=1$. In the physically interesting $z_{\text{c}}\ll1$ regime of
highly compact exotic objects \cite{Notephy}, one finds a remarkably
good agreement between the approximated radii
$\{z^{\text{analytical}}_{\text{c}}(n)\}$ of the compact objects
that can support the marginally-stable scalar resonances [as
calculated from the analytical resonance formula (\ref{Eq31})] and
the corresponding exact radii
$\{z^{\text{numerical}}_{\text{c}}(n)\}$ of the compact exotic
objects [as determined numerically from the resonance equation
(\ref{Eq23})].} \label{Table4}
\end{table}

\section{Analytical vs. former numerical results}

It is physically important to compare our {\it analytical} results
for the marginally-stable exotic compact objects with the
corresponding {\it numerical} data published recently in the very
interesting work of Maggio, Pani, and Ferrari \cite{Pan}. In Table
\ref{Table5} we display, for various values of the compactness
parameter $\delta\equiv (r^{\text{max}}_{\text{c}}-r_+)/M$
\cite{Notedx} introduced in \cite{Pan}, the dimensionless ratio
${\bar a}^{\text{analytical}}(\delta)/{\bar
a}^{\text{numerical}}(\delta)$. Here $\{{\bar
a}^{\text{analytical}}(\delta)\}$ are the analytically derived
dimensionless angular momenta which characterize the critical
(marginally-stable) spinning exotic compact objects, and $\{{\bar
a}^{\text{numerical}}(\delta)\}$ are the corresponding numerically
computed \cite{Pan} values of the critical rotation parameter.
Interestingly, from the data presented in Table \ref{Table5} one
finds a remarkably good agreement between our {\it analytical}
formulas and the corresponding {\it numerical} data of \cite{Pan}.

\begin{table}[htbp]
\centering
\begin{tabular}{|c|c|c|c|c|c|}
\hline $\delta\equiv {{(r_{\text{c}}-r_+)}/{M}}$ & \ \ \ $10^{-5}$\
\ \ \ & \ \ \ $10^{-4}$\ \ \ \ & \ \ \ $10^{-3}$\ \ \ \ & \ \ \ $10^{-2}$\ \ \ \ & \ \ \ $10^{-1}$\ \ \ \ \\
\hline \ ${{{\bar a}^{\text{analytical}}}\over{{\bar
a}^{\text{numerical}}}}$\ \ &\ $0.999$\ \ &\ $0.999$\ \
&\ $1.001$\ \ &\ $1.000$\ \ &\ $1.001$\ \ \\
\hline
\end{tabular}
\caption{Marginally-stable spinning exotic compact objects. We
display, for various values of the compactness parameter
$\delta\equiv (r^{\text{max}}_{\text{c}}-r_+)/M$ \cite{Notedx}
introduced in \cite{Pan}, the dimensionless ratio ${\bar
a}^{\text{analytical}}(\delta)/{\bar a}^{\text{numerical}}(\delta)$
between the analytically derived critical angular momentum which
characterizes the marginally-stable spinning exotic compact object
and the corresponding numerically computed \cite{Pan} value of the
critical rotation parameter. The data presented are for the
fundamental ($n=1$) resonances of the spinning exotic compact
objects with reflecting Dirichlet boundary conditions and for a
massless scalar field mode with $l=m=1$. One finds a remarkably good
agreement between our {\it analytical} results and the corresponding
{\it numerical} data of \cite{Pan}.} \label{Table5}
\end{table}

\section{Summary and Discussion}

Horizonless exotic compact objects with reflective properties have
recently attracted much attention from physicists as possible
quantum-gravity alternatives to classical black-hole spacetimes (see
\cite{eco1,eco2,eco3,eco4,eco5,eco6,eco7,eco8,eco9,eco10,eco11,eco12,Pan}
and references therein). In a physically important work, Maggio,
Pani, and Ferrari \cite{Pan} have recently studied numerically the
stability properties of a family of horizonless spinning exotic
compact objects which are characterized by curved spacetime
geometries that modify the Kerr metric only at some microscopic
scale around the would-be classical horizon.

The numerical analysis presented in \cite{Pan} has demonstrated that
horizonless spinning compact objects with reflective surfaces may
become superradiantly unstable to scalar perturbation modes
\cite{Notebi}. Intriguingly, however, the results presented in
\cite{Pan} have revealed the important fact that the ergoregion
instability shuts down if the quantum reflective surface of the
horizonless compact object is located far enough from the would-be
classical black-hole horizon. This highly interesting numerical
result implies, in particular, that there exist a unique family of
{\it marginally-stable} spinning exotic compact objects which
determine the critical boundary between stable and unstable
horizonless spinning configurations.

In the present paper we have used {\it analytical} techniques in
order to explore the physical properties of the critical
(marginally-stable) spinning exotic compact objects. These
horizonless reflecting objects are characterized by their ability to
support spatially regular static configurations of massless scalar
fields in their exterior spacetime regions. The physical
significance of this unique family of critical (marginally-stable)
exotic compact objects stems from the fact that it marks the
boundary between stable and unstable composed
spinning-exotic-compact-object-massless-scalar-field configurations.

The main results derived in the present paper and their physical
implications are as follows:

(1) We have proved that, for given angular harmonic indices $(l,m)$
of the massless scalar field mode, there exist two {\it discrete}
spectra of radii, $\{r^{\text{Dirichlet}}_{\text{c}}({\bar
a},l,m;n)\}^{n=\infty}_{n=1}$ and
$\{r^{\text{Neumann}}_{\text{c}}({\bar
a},l,m;n)\}^{n=\infty}_{n=1}$, which characterize the horizonless
spinning compact objects with reflective boundary conditions that
can support the spatially regular {\it static} (marginally-stable)
scalar field configurations \cite{Notemsm,Hodrc,Herkr}. In
particular, we have shown that the analytically derived resonance
equations [see Eqs. (\ref{Eq22}) and (\ref{Eq23})]
\begin{equation}\label{Eq32}
{_2F_1}(l+1-2i\alpha,l+1;2l+2;1-x^{\text{Dirichlet}}_{\text{c}})=0\
\ \ \ \ \text{and}\ \ \ \ \
{{d}\over{dx}}[x^{-i\alpha}(1-x)^{l+1}{_2F_1}(l+1-2i\alpha,l+1;2l+2;1-x)]_{x=x^{\text{Neumann}}_{\text{c}}}=0\
\end{equation}
determine the critical dimensionless radii of the marginally-stable
spinning exotic compact objects.

(2) We have shown that the physical properties of the critical ({\it
marginally-stable}) horizonless spinning objects can be studied {\it
analytically} in the physically interesting regime
$x_{\text{c}}\ll1$ which corresponds to highly compact exotic
objects \cite{Notephy}. In particular, using analytical techniques,
we have derived the remarkably compact resonance formula [see Eqs.
(\ref{Eq30}) and (\ref{Eq31})]
\begin{equation}\label{Eq33}
x_{\text{c}}({\bar a},l,m;n)={e^{-\pi (n+\epsilon)/\alpha}}\times
{\cal F}\ \ \ ; \ \ \ n\in\mathbb{Z}\  ,
\end{equation}
where
\begin{equation}\label{Eq34}
{\cal F}({\bar a},l,m)=
\Big[{{\Gamma(2i\alpha)\Gamma(l+1-2i\alpha)}\over{\Gamma(-2i\alpha)\Gamma(l+1+2i\alpha)}}\Big]
^{1/2i\alpha}\ \ \ \ ; \ \ \ \ \epsilon=\begin{cases}
{1\over2} &\ \ \ \ \text{Dirichlet B. C.}\  \\
0 &\ \ \ \ \text{Neumann B. C.}\ \
\end{cases}
\end{equation}
for the discrete spectra of dimensionless radii which characterize
the spinning compact objects that can support the static
(marginally-stable) massless scalar field configurations.

It is worth stressing the fact that the physical significance of the
largest (outermost) radii $r^{\text{max}}_{\text{c}}({\bar a})\equiv
\text{max}_n\{r_{\text{c}}({\bar a};n)\}$ stems from the fact that
these critical dimensionless radii mark the boundary between stable
and unstable horizonless compact objects with reflecting boundary
conditions. In particular, spinning exotic compact objects whose
reflecting surfaces are located in the radial region
$r_{\text{c}}<r^{\text{max}}_{\text{c}}({\bar a})$ are
superradiantly unstable to scalar perturbation modes, whereas
spinning exotic compact objects whose reflecting surfaces are
located in the radial region
$r_{\text{c}}>r^{\text{max}}_{\text{c}}({\bar a})$ are stable.

(3) It has been explicitly demonstrated (see Tables \ref{Table3} and
\ref{Table4}) that the analytically derived resonance spectra
(\ref{Eq30}) and (\ref{Eq31}), which determine the dimensionless
discrete radii of the horizonless exotic objects that can support
the static (marginally-stable) scalar resonances, agree in the
physically interesting regime $x_{\text{c}}\ll1$ of highly-compact
objects \cite{Notephy} with the corresponding exact radii of the
critical exotic compact objects [as determined numerically directly
from the resonance equations (\ref{Eq22}) and (\ref{Eq23})].

(4) It is worth pointing out that the compact resonance spectra
(\ref{Eq33}), which characterize the marginally-stable horizonless
configurations, can be further simplified in the slow rotation
regime ${\bar a}\ll1$. In particular, one finds from (\ref{Eq33})
and (\ref{Eq34}) the resonance spectra \cite{Notesma}
\begin{equation}\label{Eq35}
x_{\text{c}}({\bar a}\ll1,l,m;n)={e^{-2\pi
(n+\epsilon-{1\over2})/m{\bar a}}}\times e^{-2[\psi(l+1)-\psi(1)]}\
\ \ ; \ \ \ n\in\mathbb{Z}\
\end{equation}
in the regime ${\bar a}\ll1$ [or equivalently, $\alpha\simeq m{\bar
a}/2\ll1$, see Eq. (\ref{Eq14})] of slowly-rotating exotic compact
objects. It is worth noting that this analytical formula is
especially useful since it is highly difficult to probe numerically
\cite{Pan} the small ${\bar a}\ll1$ regime [which, as is evident
from (\ref{Eq35}), corresponds to exponentially small values of the
dimensionless radius $x_{\text{c}}({\bar a}\ll1)$].

(5) Finally, we have explicitly demonstrated that our {\it
analytical} formulas agree remarkably well with the corresponding
{\it numerical} data that recently appeared in the interesting work
of Maggio, Pani, and Ferrari \cite{Pan}.

\bigskip
\noindent
{\bf ACKNOWLEDGMENTS}
\bigskip

This research is supported by the Carmel Science Foundation. I thank
Yael Oren, Arbel M. Ongo, Ayelet B. Lata, and Alona B. Tea for
stimulating discussions.

%\newpage


\begin{thebibliography}{99}
\bibitem{Notebi} It is worth noting that the superradiant
instability phenomenon observed in \cite{Bri} is also expected to
characterize the dynamics of other bosonic (integer-spin) fields in
the curved spacetimes of these horizonless spinning exotic compact
objects.
\bibitem{Notemsm} It is worth mentioning that spinning Kerr black holes, as opposed
to the horizonless spinning compact objects that we study in the
present paper, cannot support static spatially regular massless
scalar fields in their exterior regions. However, as recently
demonstrated in \cite{Hodrc,Herkr}, spinning black holes can support
stationary (rather than static) massive scalar field configurations.
\bibitem{Noteun} We shall use natural units in which $G=c=\hbar=1$.
\bibitem{Notebl} Here we have used the familiar Boyer-Lindquist coordinates $(t,r,\theta,\phi)$.
\bibitem{Noteman} Following the physically interesting model of the exotic compact
objects studied in \cite{Pan}, we shall assume that the energy and
angular momentum of the reflective surface are negligible.
\bibitem{Notesk} It is important to stress the fact that the assumption
made in the physical model of \cite{Pan}, according to which the
exterior curved spacetimes of the spinning exotic objects are
described by the familiar Kerr line element (\ref{Eq1}), is a
non-trivial one. As discussed in \cite{Pan} (see also
\cite{kr1,kr2,kr3}), this assumption is expected to be valid in the
physically interesting regime $z_{\text{c}}\ll1$ of exotic compact
objects whose reflecting quantum membranes are located a microscopic
distance above the would-be classical horizons.
\bibitem{Notedec} The physical parameters $\{\omega,l,m\}$ are
respectively the proper frequency, the spheroidal harmonic index,
and the azimuthal harmonic index (with $-l\leq m\leq l$) which
characterize the massless scalar field mode.
\bibitem{Noteby} For brevity, we shall henceforth omit the angular
harmonic indices $(l,m)$ of the massless scalar field mode.
\bibitem{Noteklm0} Here we have used the relation $K_{lm}(a\omega=0)=l(l+1)$ [see Eq. (\ref{Eq8})].
\bibitem{Notezx} Note that the dimensionless coordinate
$z_{\text{c}}$, which characterizes the radius of the horizonless
reflecting exotic compact object, can be expressed in the form
$z_{\text{c}}={{x_{\text{c}}}\over{1-x_{\text{c}}}}\cdot{{r_+-r_-}\over{r_+}}$
[see Eqs. (\ref{Eq14}) and (\ref{Eq25})].
\bibitem{Notenn} Here we have used the relation $1=e^{-2i\pi n}$, where $n$ is an integer.
This dimensionless resonance parameter characterizes the composed
spinning-exotic-compact-object-massless-scalar-field configurations.
\bibitem{Notegam1} Taking cognizance of Eq. 6.1.23 of
\cite{Abram}, one finds the relations
$\Gamma(2i\alpha)/\Gamma(-2i\alpha)=e^{i\phi_1}$ and
$\Gamma(l+1-2i\alpha)/\Gamma(l+1+2i\alpha)=e^{i\phi_2}$, where
$\{\phi_1,\phi_2\}\in\mathbb{R}$. These characteristic relations
imply, in particular, that
$\{x^{\text{Dirichlet}}_{\text{c}}(n),x^{\text{Neumann}}_{\text{c}}(n)\}\in\mathbb{R}$.
\bibitem{Notephy} As emphasized above, the small-$x_{\text{c}}$ regime (\ref{Eq26})
corresponds to the physically interesting family, that was recently
studied numerically in \cite{Pan} (see also
\cite{eco1,eco2,eco3,eco4,eco5,eco6,eco7,eco8,eco9,eco10,eco11,eco12}),
of highly compact horizonless exotic objects whose quantum
reflective surfaces are located very near the would-be classical
black-hole horizons [see Eqs. (\ref{Eq2}) and (\ref{Eq3})].
\bibitem{Notedx} Note that the dimensionless coordinate
$\delta$ introduced in \cite{Pan}, which characterizes the radius of
the horizonless reflecting exotic compact object, can be expressed
in the form
$\delta={{x_{\text{c}}}\over{1-x_{\text{c}}}}\cdot{{r_+-r_-}\over{M}}$
[see Eq. (\ref{Eq14})].
\bibitem{Notesma} Here we have used the relations
$[\Gamma(2i\alpha)/\Gamma(-2i\alpha)]^{1/2i\alpha}=e^{2\psi(1)+\pi/2\alpha}[1+O(\alpha)]$
and
$[\Gamma(l+1-2i\alpha)/\Gamma(l+1+2i\alpha)]^{1/2i\alpha}=e^{-2\psi(l+1)}[1+O(\alpha)]$
in the $\alpha=(m{\bar a}/2)\cdot[1+O({\bar a}^2)]\ll1$ regime,
where $\psi(z)$ is the Psi (Digamma) function (see Eq. 6.3.1 of
\cite{Abram}).
\end{thebibliography}
\end{document}